\documentclass[prb, preprint, letter, superscriptaddress]{revtex4-1}
\usepackage{amssymb}
\usepackage{amsmath}
\usepackage{bbm}
\usepackage{graphicx}
\usepackage{color}

\newcommand{\ris}{\{\vec r_i\}}
\newcommand{\rips}{\{\vec{r'}_i\}}
\newcommand{\Lne}{\hat L_{_\textrm{NE}}}

\newcommand{\rip}{\vec{r'}_i}
\newcommand{\rjp}{\vec{r'}_j}
\newcommand{\ri}{\vec{r}_i}
\newcommand{\rj}{\vec{r}_j}

\begin{document}

\author{Shenshen Wang}
\affiliation{Department of Chemical Engineering, Department of Physics, Massachusetts Institute of Technology, Cambridge, MA 02139, USA}
\author{Peter G. Wolynes}
%\affiliation{Department of Physics, Department of Chemistry and Biochemistry,  and Center for Theoretical Biological Physics, University of California, San Diego, La Jolla, CA 92093, USA}
\affiliation{Department of Chemistry, and Center for Theoretical Biological Physics, Rice University, Houston, TX 77005, USA}

\title{Active patterning and asymmetric transport in a model actomyosin network}
\date{\today}

\begin{abstract}
Cytoskeletal networks, which are essentially motor-filament assemblies, play a major role in many developmental processes involving structural remodeling and shape changes. These are achieved by nonequilibrium self-organization processes that generate functional patterns and drive intracellular transport. We construct a minimal physical model that incorporates the coupling between nonlinear elastic responses of individual filaments and force-dependent motor action. By performing stochastic simulations we show that the interplay of motor processes, described as driving anti-correlated motion of the network vertices, and the network connectivity, which determines the percolation character of the structure, can indeed capture the dynamical and structural cooperativity which gives rise to diverse patterns observed experimentally. The buckling instability of individual filaments is found to play a key role in localizing collapse events due to local force imbalance. Motor-driven buckling-induced node aggregation provides a dynamic mechanism that stabilizes the two dimensional patterns below the apparent static percolation limit. Coordinated motor action is also shown to suppress random thermal noise  on large time scales, the two dimensional configuration that the system starts with thus remaining planar during the structural development. By carrying out similar simulations on a three dimensional anchored network, we find that the myosin-driven isotropic contraction of a well-connected actin network, when combined with mechanical anchoring that confers directionality to the collective motion, may represent a novel mechanism of intracellular transport, as revealed by chromosome translocation in the starfish oocyte.
\end{abstract}

\hyphenation{}

\maketitle

\section{Introduction}

%\textbf{active network structure: motor-filament assembly (force generators mixed with structural elements)}

The cytoskeleton is a complex meshwork of biopolymers spanning the cytoplasm of eukaryotic cells. It consists of filamentous proteins that are organized into higher-order assemblies by a myriad of auxiliary proteins such as crosslinkers and bundlers, as well as capping and severing proteins, which bind to the scaffolding elements. The regulation of the kinetics of these binding proteins, the polymerization and depolymerization processes of the filaments and the action of molecular motors vivify the cytoskeleton, changing it from a static carcass into a dynamic and versatile muscle.\cite{cytoskeleton book chap1}

Cytoskeletal networks play a major role in many developmental processes involving structural remodeling and shape changes, which range from cytokinesis and cell motility to wound healing and tissue morphogenesis. Unlike macroscopic machines, within the active cellular materials there is no clear distinction between the force generators and the structural elements. Rather, the force-generating motor proteins are intimately mixed with the elementary building blocks of cell structure on a molecular scale.
An important motor-filament assembly---the kinesin-microtubule system---forms well-focused mitotic spindle poles in order to accomplish high-accuracy segregation of replicated chromosomes into daughter cells. Networks of filamentous actin (F-actin) and the type-II myosin motors have been identified as the major components of the cellular contractile machinery. Walking on the structural scaffold provided by an actin network, myosin-II motors themselves self-assemble into bipolar minifilaments \cite{minifilaments} that generate sustained sliding of neighboring actin filaments relative to each other. By carrying out such correlated motions the minifilaments reorganize the actin networks and generate tension ultimately powered by ATP hydrolysis. The formation and coalescence of actomyosin condensates to exert contractile forces have been seen in contractile rings driving cytokinesis \cite{cytokinesis 1, cytokinesis 2} and wound healing, \cite{wound healing} and in contractile networks that deform epithelial cell layers in developing embryos \cite{developing embryos 1, developing embryos 2, contractile ratchet} and drive polarizing cortical flows.\cite{polarizing cortical flows 1, polarizing cortical flows 2}

%- \textbf{mechanical: nonlinearity (relieve compressive stress through filament buckling; localize damage)}

%Physical studies concentrating on the mechanical properties of cells have been useful in elucidating the \emph{synergies} of generic physical mechanisms and specific biological regulation in establishing the overall mechanical behavior of biological cells.

%A remarkable universality has been revealed in the viscoelastic responses of reconstituted networks and of whole cells of different types over a wide range of timescales, reminiscent of soft glassy materials \cite{soft glassy 1, soft glassy 2, soft glassy 3}.

%On the other hand, functionally relevant heterogeneities exemplify the specificity of biochemical signaling events. One remarkable demonstration is provided by the formation of filopodia (finger-like extensions of cytoplasm sent out by the cell in motion at its leading edge), in which the local activation of various actin-binding proteins enables the dynamic generation of highly localized meso-structures \cite{meso structure 1, meso structure 2}.

The unusual mechanical properties of individual filaments have been highlighted especially as to their strain-stiffening \cite{nonlinear stiffening} behavior in response to a sustained stretching, and the negative normal stresses \cite{negative normal stress chap1} under shear reflecting the buckling instability \cite{reversible stress softening} when subjected to a compressive load.
A multistage aggregation process \cite{multistage coarsening} observed in a reconstituted system of purified actin and myosin suggests that motor-induced buckling in connected actin structures underlies the self-organized contractile state and its dynamics.
Recent experiments on a miniature of the cell cortex \cite{buckling coordinates contractility and severing} have further shown that over a wide range of conditions, the extent of network contraction corresponds precisely to the extent of individual F-actin shortening via buckling. This supports the essential role of filament buckling in breaking the symmetry of the response to tensile versus compressive forces in order to facilitate large-scale contraction of the actomyosin network.
Such nonlinearity in elastic responses is also found to be crucial to keep biological networks, such as spider webs, functioning mechanically in tolerance to a large number of local defects in the network structure. \cite{spider web} This flaw-tolerant feature relies on the existence of a softening regime that allows localization of the damage.

%- \textbf{dynamical: contractility and flow--local force imbalance captured by mechano-sensitive motor response (ATP dependent)}

%\emph{$\rightarrow$ coupling between these two aspects: nonequilibrium self-organization of motor-filament assemblies that generates functional patterns and allows intracellular transport}

%The formation and dynamics of contractile structures are manifested in pulsed contractions of an actomyosin network that drive epithelial sheet deformation during morphogenesis \cite{contractile ratchet chap1}. Such actomyosin aggregates also are responsible for a multistage coarsening process that occurs in a bottom-up model system for contractility \cite{multistage coarsening chap1}.

Cytoskeletal networks self-organize into highly dynamic and heterogeneous functional patterns from the interplay between active force generation by molecular motors and passive dissipation of energy in the crowded cellular interior.
In reconstituted filament-motor assemblies there arise some comparatively regular patterns such as asters, \cite{polarity sorting 0, actin aster, active gels} in which stiff filaments or filament bundles radiate from a common center. These asters resemble the mitotic spindles formed in dividing cells.
Formation of such ordered patterns is driven by a polarity sorting mechanism \cite{actin aster, polarity sorting 1 chap1} and is well captured by the continuum theories of active polar gels \cite{active polar gels 1, active polar gels 2, active polar gels 3} based on symmetry considerations.
On the other hand, irregular heterogeneous cluster structures have also been seen both in the actomyosin networks of \textit{C. elegans} embryos \cite{asymmetrical contraction chap1} and in the minimal in vitro network model \cite{steady clusters chap1}.
In these quasi-two-dimensional systems, the regular structures and disordered aggregates observed in vitro and in vivo seem to reveal various self-organization schemes. Many theoretical attempts have been made to uncover the underlying physical mechanisms. \cite{self organization theory 1, self organization theory 2, self organization theory 3, self organization theory 4, Chase_pull out soft, Carlsson_contractility}
Recently we have constructed a minimal physical model of the actomyosin network and studied the formation and dynamics of the contractile structures using stochastic simulations.\cite{contractility}
Here we generalize our original three dimensional study to two dimension, in close resemblance to the largely two dimensional character of the biological systems discussed above. We will show that, starting with a sufficiently connected two dimensional structure, the interplay of buckling instability and motor susceptibility to imposed mechanical forces allows for the emergence of both regular and heterogeneous patterns.
In contrast to striated muscle contraction which crucially depends on the highly organized sarcomeric structure, \cite{Hill_38_49} onset of contractility and the occurrence of large-scale force heterogeneity in disordered actomyosin networks arise from the nonlinear elastic response of individual actin filaments and the force-dependent motor action which introduces inhomogeneities in motor speeds. A similar mechanism has previously been invoked \cite{bundle buckling} to explain bundle contractility in the absence of sarcomeric arrangements. \cite{non-sarcomeric bundle 1-3}
Interestingly, motor-driven buckling-induced node aggregation extends the apparent static percolation threshold to lower connectivity and allows the formation of stable patterns composed of node aggregates connected by tense filaments, even below the isostatic point where the shear modulus of a static elastic network vanishes.\cite{isostaticity 1, isostaticity 2} This dynamic mechanism bears some resemblance to another proposal \cite{isostaticity 2} that however gives an equilibrium origin for the shifted isostatic point.

In addition to the formation of quasi-steady functional patterns, non-steady collective directed motion driven by active processes is just as crucial to biology, if not more so. Such motion arises from ATP-dependent nonequilibrium processes that couple the movements of the numerous constituents to their locations in the structure. Cytoplasmic streaming, \cite{cytoplasmic streaming chap1} a cell-scale coherent flow mainly driven by kinesin-microtubule assemblies, \cite{kinesin driven streaming 1, kinesin driven streaming 2} dramatically illustrates the nonequilibrium nature of cellular dynamics and the collective motion of a myriad of constituents.
Apart from the relatively random mixing of cellular contents via the streaming flow, other mechanisms are needed in order to enable directed transport of matter and information to distinct subcellular locations.
Recent experimental studies \cite{intracellular transport} of F-actin driven chromosome transport in starfish oocyte have suggested a novel mechanism of intracellular transport, in which a 3D F-actin meshwork contracts homogeneously and isotropically throughout the nuclear space. Mechanical anchoring of the network to the cell cortex has been shown to confer directionality to this large-scale motion. However, the experimental attempts to test the involvement of myosin motors in this process remains inconclusive.
By performing stochastic simulations on an anchored 3D amorphous network, where anti-correlated node motion mimicking myosin-driven sliding of neighboring actin filaments distributes homogeneously over the structure, we have indeed observed an isotropic contraction of the actin network toward its center of mass, which approaches the mechanical anchor at a constant speed, as seen experimentally.
This finding supports the key role of myosin motors in correlating local movements and their interplay with network architecture to coordinate the collective motion. Moreover, it suggests an intriguing possibility that the contractile subunits similar in composition and organization to those that form the 2D meshwork under the cell membrane, to mediate cytokinesis or morphogenesis, may alternatively organize into 3D networks to drive intracellular transport.

\section{Model}

%\textbf{$\ast$ our model}

%\textbf{- structure: CC, motors and kicks, quenched distribution of motors over bonds}

%The ``cat's cradle" model \cite{cats} was introduced by Shen and Wolynes to study the statistical mechanics of a collection of buckling bonds connecting point nodes sitting on a regular lattice.

As in our earlier studies,\cite{equil_CSK, flow, contractility} the cytoskeleton is treated as a crosslinked network of nonlinear elastic filaments. The nonlinear elasticity of individual filaments allows them to stiffen under strain (with an effective elastic modulus $\beta\gamma$) thereby preventing large deformations that could threaten tissue integrity. On the other hand, the buckling instability (when shortened below the relaxed length $L_e$) occurs to relieve the compressive load and is shown to be crucial for large-scale contraction \cite{multistage coarsening, buckling coordinates contractility and severing} that makes possible wound healing and morphogenesis.
The role of filament buckling in contraction has been discussed for one-dimension actomyosin bundles \cite{bundle buckling}. We will consider more realistic two dimensional and three dimensional disordered actomyosin networks and show how the local buckling events again lead to a global patterning.
Note that what is essential for contraction is not necessarily the buckling instability per se, but rather the nonlinearity in elastic responses, in particular, the asymmetric response to stretching versus compressive loads---an inherent property of the actin filaments. A recent simulation study of a 2D random actomyosin network \cite{Carlsson_contractility} has highlighted an alternative realization of nonlinear elastic response via bending or rotational modes that lead to low-energy contractile configurations.

Rather than keep track of the filament degrees of freedom based on the contour of the filaments using a continuum description, we follow the motion of the vertices or the nodes of the network. This model encodes the effect of the filaments (i.e. their asymmetric load response) via the pairwise interaction between the bonded nodes. The cat's cradle type potential energy \cite{cats, equil_CSK} is given by $U(r)=(1/2)\beta\gamma(r-L_e)^2\Theta(r-L_e)$, where $\Theta(\cdot)$ is the Heaviside step function. Elastic energy only arises when the contour length $r$ between the nodes exceeds the relaxed length $L_e$ of individual filaments. We will assume vanishing excluded volume effects to allow large-scale structural rearrangement and compact node aggregation.

\begin{figure}[htb]
\centerline{\includegraphics[angle=0, scale=0.45]{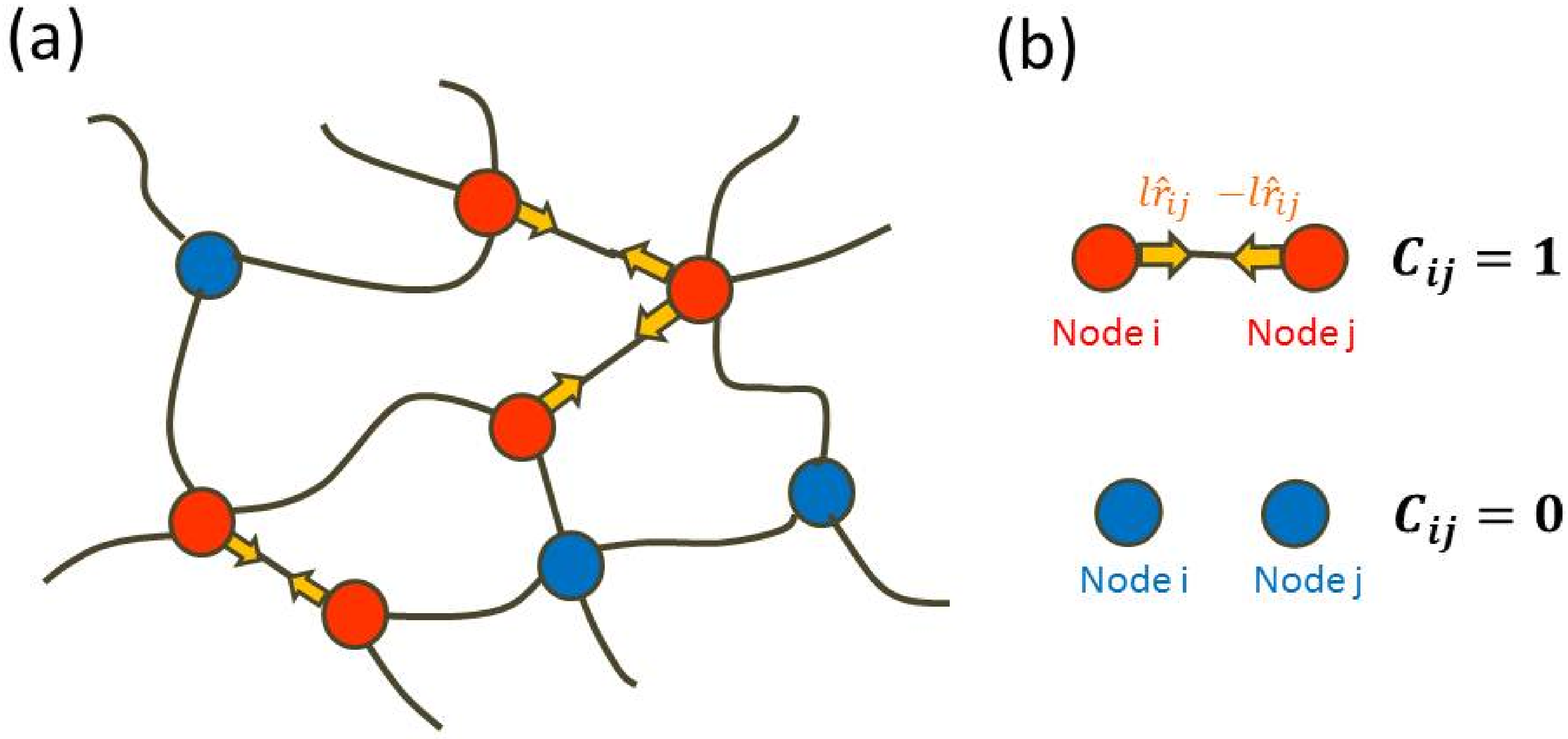}}
\caption{Schematic of the model.
(a) Model actomyosin network consisting of actin filaments (lines) connected by crosslinks (circles). The crosslinks or nodes can be passive (blue), i.e. all the bonds connected to it bear no myosin motors, or active (red), i.e. they are subjected to anti-correlated motor action (orange arrows) mimicking myosin-driven relative sliding of neighboring actin filaments that effectively pulls together the node pairs. If all the bonds are motor-attached, then passive nodes have no bonds at all. (b) A pair of nodes (i and j) either move under motor kicks with step size $l$ toward each other ($C_{ij}=1; \vec{l}_{ij}=-\vec{l}_{ji}=l\hat{r_{ij}}$), or remain non-bonded ($C_{ij}=0$) and only move under thermal agitation.
}
\label{model}
\end{figure}

Motors do not explicitly enter the model; instead, they are exemplified as generating kicks on the motorized nodes (Fig.~\ref{model}).
To mimic the motor-driven filament sliding in actomyosin networks, as before,\cite{contractility, effective interaction} the motors are described as generating anti-correlated kicks on pairs of actively-bonded nodes that pull in slack locally.
A kick pair acting on nodes $i$ and $j$ can be represented by a pair of displacements along the line of centers $(\vec l_{ij},\vec l_{j\,i})=l(\hat r_{ij},-\hat r_{ij})$, where $l$ is the fixed kick step size and $\hat r_{ij}$ is a unit vector pointing from node $i$ to node $j$.
The dynamical evolution of the many-particle configuration $\ris$ due to these motor-driven nonequilibrium (NE) processes can be described by a master equation $\partial\Psi/\partial t=\Lne\Psi$ for the probability distribution function $\Psi(\ris;t)$ with $\Lne\Psi(\ris;t)=\int\Pi_id\vec{r'}_i\left[K(\rips\rightarrow\ris)\Psi(\rips;t)-K(\ris\rightarrow\rips)\Psi(\ris;t)\right],$
where the integral kernel $K(\rips\rightarrow\ris)$ encodes the probability of transitions between different node configurations.
The motor kicks occur at a load-dependent rate given by
$k=\kappa[\Theta(\Delta U)\exp(-s_u\beta\Delta U)+\Theta(-\Delta U)\exp(-s_d\beta\Delta U)]$, %\label{model rate}
where $\kappa$ is the basal kicking rate and $s_u$($s_d$) denotes motor susceptibility to energetically uphill (downhill) moves.
The free energy change $\Delta U$ is due to \textit{pairs} of node displacements.
Assuming symmetric motor susceptibility, i.e. $s_u=s_d=s$, we write explicitly
\begin{eqnarray}\label{Lne}
\nonumber&&\Lne\Psi(\ris;t)=\frac{1}{2}\kappa\sum_i\sum_j C_{ij}\\ \nonumber
&&\times \Big\{e^{-s\beta\left[U(\ri,\rj)-U(\ri-\vec l_{ij},\rj+\vec l_{ij})\right]}
\Psi(\{\cdots,\rip=\ri-\vec l_{ij},\cdots,\rjp=\rj+\vec l_{ij},\cdots\};t) \\
&& -e^{-s\beta\left[U(\ri+\vec l_{ij},\rj-\vec l_{ij})-U(\ri,\rj)\right]}\Psi(\{\cdots,\ri,\cdots,\rj,\cdots\};t)\Big\}.
\end{eqnarray}
The quantity $C_{ij}$, much like an element of the contact maps used in the description of protein structures, defines whether the node pair ($i,j$) is connected by an active bond and thus subject to anti-correlated displacements ($\vec l_{ij}, -\vec l_{ij}$) with $C_{ij}=C_{ji}=1$, or whether it remains non-bonded with $C_{ij}=C_{ji}=0$. Here we assume a high concentration of motors and abundant ATP such that all the bonds are motor-attached.

We will assume that kicks on different pairs of nodes at any given time are uncorrelated.
The rates of possible kicking events depend on the \textit{instantaneous} node configuration reflecting an assumed Markovian character of the motor dynamics.
Note that the motor power strokes and thus the kick steps are discrete occurring in a stochastic fashion. The correlated motions collapse bonds locally while they simultaneously pull taut neighboring filaments. Such correlated motions proceed until either a global balance is reached or a macroscopic collapse occurs, depending on whether the motors are susceptible and thus downhill prone in their motion (with a large positive $s$) or load-resisting (with a small or negative $s$), respectively, as well as on the network connectivity.
The sign of $s$ can be inferred on the single molecule level from the force-lifetime curve obtained by pulling experiments on single motor proteins. \cite{Alon, Dudko} Increase (Decrease) in lifetime under an increasing load would indicate a positive (negative) susceptibility, corresponding to ``catch bond" (``slip bond") behavior observed experimentally. \cite{Alon, Guo}

In this microscopic model the network connectivity and the motor distribution over the bonds of the network are quenched once they are initially assigned, so that the nonequilibrium dynamics and structures predicted by the model arise solely from the intrinsic activity of motors firmly built into the network driving correlated motions stochastically.
This assumption is in line with the fact that the structures studied in vitro are irreversibly assembled because many of the severing/capping protein factors found in vivo that allow fast pattern renewals are left out of the reconstitution. Thus disassembly of contractile structures and the transient action of crosslinking proteins, while important in vivo, are mostly absent in the simplified in vitro systems.

Our model highlights the key role played by the motor susceptibility, $s$, a parameter that characterizes how sensitively the motors respond to imposed forces.
The coupling between motor kinetics and the structure leads to a double-way feedback: Motor action induces structural changes of the network; these changes in turn modify the local mechanical environment of the motors, again, changing the load-dependent motor response.

\section{Results}

\subsection{Active patterning in quasi-2D systems}

%\textbf{- describe simulation procedure, key and new ingredients}

To study the dynamic development of quasi-2D patterns, we start from a 2D configuration with $N=225$ nodes residing on a square lattice (lattice spacing is taken to be the length unit). Periodic boundary conditions are applied. We randomly assign bonds between nearest-neighbor node pairs with a probability $P_c$, the network connectivity. Thus, structural disorder is inherent in the randomness of bond connectivity as long as $P_c<1$ (see Fig.~\ref{nonpercolating}a for an example).
While the connectivity is strictly two dimensional, we then allow the nodes to move in 3D. On time scales shorter than the time interval between two successive motor kicks ($\propto\kappa^{-1}$), dynamics is governed by diffusive relaxation at thermal temperature $T=1/(k_B\beta)$. We thus simulate the non-equilibrium dynamics described by the master equation (Eq.~\ref{Lne}) using Gillespie algorithm,\cite{Gillespie} and make intermediate thermal moves according to Langevin dynamics in the overdamped limit described by $\Delta\vec{r}_i(t)=\beta D_0\vec{F}_i\Delta t+\vec{\eta}_i(\Delta t)$, i.e., node $i$ moves by $\Delta\vec{r}_i$ during time $\Delta t$ in response to a deterministic mechanical force $\vec{F}_i=-\nabla_i U$ and the Gaussian white noise $\vec{\eta}_i$ with $\langle\eta_{i\alpha}\rangle=0$ and $\langle\eta_{i\alpha}(t)\eta_{j\beta}(t')\rangle=2D_0\delta(t-t')\delta_{ij}\delta_{\alpha\beta}$. Here $D_0$ is the thermal diffusion constant. Energies are measured in units of $k_B T$. We assume the existence of a high concentration of motor proteins and a surplus of chemical fuel such that all the bonds are motor-attached, i.e., $C_{ij}=1$ for any bonded node pair. Therefore, active nodes, i.e. those with motor-attached bonds, move both under motor kicks and through passive diffusion, whereas passive nodes having no bonds at all are solely subject to random thermal buffet.
Reaction channels are described as pairwise moves of motor-bonded nodes \emph{toward} each other along their line of centers, by an amount of $l$ for each node (Fig.~\ref{model}b). Implemented as such, motor-induced moves tend to collapse local bonds while stretching neighboring bonds.

Typical values for the biophysical parameters in the model include the motor kicking rate $\kappa=10\mathrm{s}^{-1}$ and the motor step size $l=50\mathrm{nm}$, consistent with the motor speed of several hundred nanometers per second as indicated experimentally. \cite{cytoskeleton book chap1} The motor kick size $l$ is much smaller than the crosslink separation $R_0$ in a well-connected network where $R_0$ is on the order of $1$--$10\mathrm{\mu m}$. We take $R_0$ as the length unit and set the time unit to be $10\mathrm{ms}$, then the model parameters used in the simulations are given by $\kappa=0.1$ and $l=0.05$ in these units.

%Upon each active move, reaction rates of all possible channels are compared to determine when and which move to occur next.

\subsubsection{Non-percolating structures}

\begin{figure}[htb]
\centerline{\includegraphics[angle=0, scale=0.48]{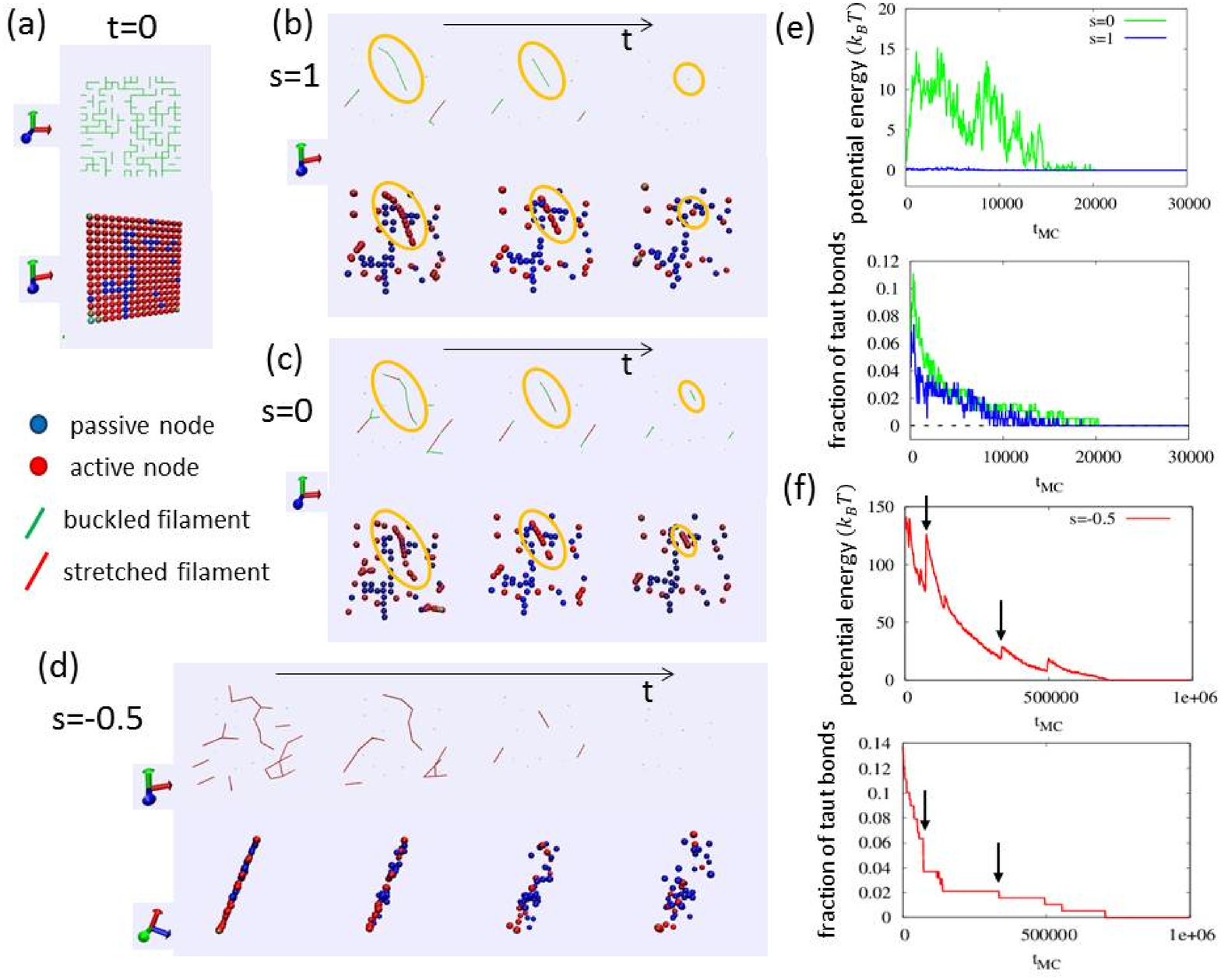}}
\caption{Structural development of non-percolating systems driven by motors with various susceptibilities.
(a) Initial bond structure (upper) and node configuration (lower) where voids in the network reflect the locations of passive nodes (blue spheres) which have no bonds at all. The initial network consists of buckled filaments (green segments) only. (b-d) Snapshots during the structural development until complete collapse of separate clusters occurs (orange circles) for $s=1$ (b), $s=0$ (c) and $s=-0.5$ (d). View angles are indicated by the axes. Lower row of (d) shows the side view of the node configuration which becomes progressively more spread out as time advances. (e-f) Time courses of the potential energy (upper) and the corresponding fraction of taut bonds (lower) for $s=1,0$ (e) and $s=-0.5$ (f). Arrows in (f) correspond to the middle two panels in (d). Note that the system with $s=-0.5$ evolves on a much longer time scale. The model parameters are $L_e=1.2, \beta\gamma=5, l=0.05, \kappa=0.1$ and $P_c=0.4$.}
\label{nonpercolating}
\end{figure}

We first look at non-percolating structures ($P_c=0.4$, i.e. mean coordination number $z<2$) where the voids in the network reflect the locations of the passive nodes (Fig.~\ref{nonpercolating}a). Non-bonded passive nodes (blue spheres in Fig.~\ref {nonpercolating}b-d) move slowly under random thermal buffets; the resulting off-plane motion leads to the diffuse node configurations at long times (side view in Fig.~\ref{nonpercolating}d lower row). In contrast the motor-bonded active nodes (red spheres) form separate clusters that rapidly collapse into compact aggregates under anti-correlated kicks (see local contraction events marked by orange circles in Fig.~\ref{nonpercolating}b-c).
Since active processes only can make their effects through bonded interactions, individual aggregates of active nodes, internally connected by buckled bonds, bear no external mechanical load and thus diffuse randomly just as the passive nodes do. Therefore, the initial planar configurations become increasingly spread out as more bonds collapse, forming a larger number of floppy clumps that diffuse aimlessly around (Fig.~\ref{nonpercolating}d).

Non-percolating systems with various motor susceptibilities exhibit distinct dynamic behaviors and evolve on different timescales.
When driven by adamant motors with $s=0$ which are insensitive to imposed mechanical forces, the system exhibits, compared to the case for susceptible motors with $s=1$, a larger number of taut bonds which are stretched more strongly. This leads to a much higher energy cost during the transients (Fig.~\ref{nonpercolating}e), though the isolated tense filaments do finally slacken (Fig.~\ref{nonpercolating}c).
For systems driven by load-resisting motors with $s=-0.5$ (Fig.~\ref{nonpercolating}d), downward jumps in the number of taut bonds are accompanied by uprising jerks in the potential energy curve (arrows in Fig.~\ref{nonpercolating}f).
These cytoskeletal ``earth quakes" happen because, as local moderately stretched bonds become buckled, the neighboring tense bonds become even more stretched. Hence even though the total number of taut bonds decreases, a higher energy cost is incurred.

Despite different transient behavior, non-percolating systems driven by motors with various $s$ values all end up completely collapsing into clumps of buckled filaments. Similarity in the steady-state structure (Fig.~\ref{nonpercolating}b-d final snapshots) is reflected in that of the statistical measures at long times (Fig.~\ref{nonpercolating}e-f); the simultaneous vanishing of potential energy and the fraction of taut bonds characterizes structures composed of disconnected floppy clusters.

\subsubsection{Effectively percolating networks}

Dramatically different behavior emerges in well-connected networks ($P_c=0.8$, thus $3<z<4$).
A sufficient connectivity allows spatial coordination between local motor-driven events, leading to $s$-dependent active patterning.
Note that the qualitative features of the observed phenomena mainly depend on the sign of $s$; varying the amplitude of $s$ merely modifies the quantitative aspects such as the time scale of structural rearrangements.

\begin{figure}[htb]
\centerline{\includegraphics[angle=0, scale=0.48]{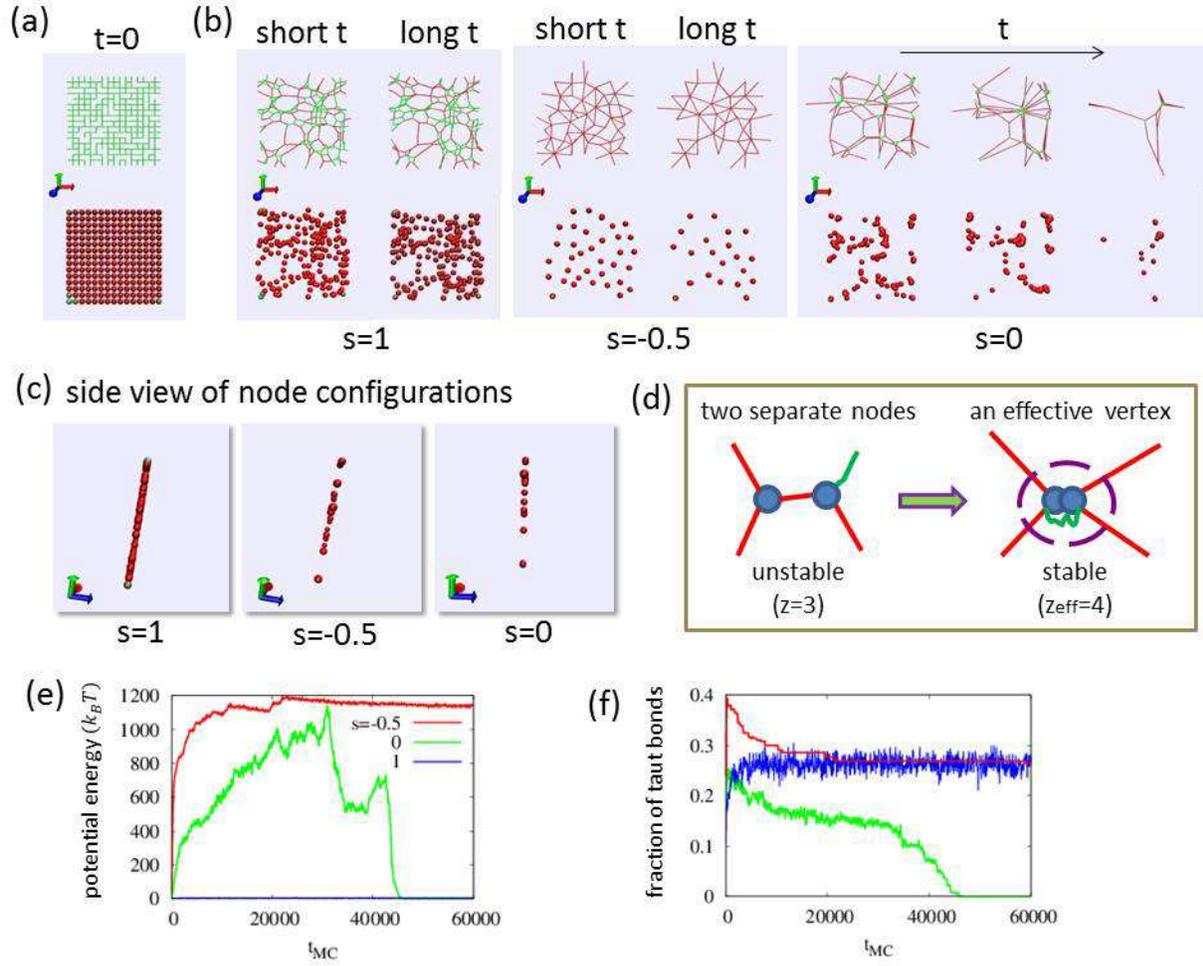}}
\caption{Active patterning of effectively percolating networks.
(a) Initial bond structure (upper) and node configuration (lower). (b) Snapshots during dynamic evolution for $s=1, -0.5$ and $0$ (left to right). The rightmost panel for $s=0$ shows the unstable structures before complete collapse occurs. (c) Side view of the node configurations which remain planar during the evolution regardless of $s$ value. (d) Schematic of the buckling-induced node aggregation that yields $z_\textrm{eff}=z_c$, given $z<z_c$. (e-f) Time courses of the potential energy (e) and the corresponding fraction of taut bonds (f). The model parameters are identical to those in Fig.~\ref{nonpercolating} except that $P_c=0.8$.}
\label{percolating}
\end{figure}

Susceptible motors ($s>0$) tend to minimize bond stretching and thus promote the formation of loose clusters inter-connected by slightly stretched filaments (Fig.~\ref{percolating}b left). These clusters consist of moderately buckled bonds, indicating that the ``defects" resulting from locally unbalanced forces are localized through moderate filament buckling (green floppy clumps in upper panels).
These heterogeneous cluster structures are formed in a globally stable state as a result of arrested phase separation \cite{arrested phase separation 1, arrested phase separation 2}. They resemble the steady structures formed in short-range attractive colloidal systems \cite{arrested phase separation 1, arrested phase separation 2} and in Lennard-Jones networks \cite{2D LJ}, yet arise from a different physical origin: The anti-correlation between motor-driven movements in our dynamic model produces an effective short-range attraction even in the buckling regime, and also leads to an enhanced long-range attraction or repulsion in a way which depends on motor susceptibility. \cite{effective interaction}
These effective interactions due to nonequilibrium motor processes underlie the diverse patterns seen in active cytoskeletal networks.

In contrast, load-resisting motors ($s<0$) tend to go against the energy gradient and thus bias the moves so as to induce strong stretching of neighboring bonds. This yields a highly tense network where the vertices, composed of compact aggregates of nodes bonded by completely buckled filaments, are arranged on a triangular-like regular lattice (Fig.~\ref{percolating}b middle).
The dramatic difference between the steady-state structures for networks driven by susceptible ($s=1$) motors and those with load-resisting ($s=-0.5$) motors is reflected in their extremely different energy values (Fig.~\ref{percolating}e); loose versus tight aggregation of nodes leads to modest versus strong bond stretching in between the aggregates, although a similar number of taut bonds is present in both systems (similar mean values for the red and blue curves in Fig.~\ref{percolating}f).
Moreover, susceptible motors allow considerable fluctuations about the mean configuration (flexibility due to buckled bonds), whereas load-resisting motors give rise to highly localized motions and thus a nearly frozen structure.
In contrast, despite having the same connectivity, networks in which there is no mechano-sensitivity ($s=0$) would experience failure of local force balance, which leads to a complete collapse (not shown) of the connected yet unstable structure that coarsens over time (Fig.~\ref{percolating}b right).
As a result, both the fraction of taut bonds and the potential energy drop to zero at long times (green curves in Fig.~\ref{percolating}e-f).

Distinct from the diffuse configurations displayed by the non-percolating structures, well-connected networks remain planar irrespective of $s$ values (Fig.~\ref{percolating}c).
Motor-driven anti-correlated moves occur along the lines of centers which lie within the starting plane, hence active motion would not cause any off-plane force component. Meanwhile, motor processes dominate over thermal noise as long as $\kappa l^2>D_0$, which is valid under physiological conditions where passive diffusion ($D_0$) is very slow within a crosslinked network immersed in the viscous cytoplasm.
Therefore, active moves effectively suppress random fluctuations, on timescales much longer than the time interval between successive motor kicks.

  %$z_{eff}$ and effective percolating network

Another appealing observation is that networks with non-zero $s$ values can form stable patterns with a mean coordination number $z$ lower than the isostatic value $z_c=2d$, where $d$ is the spatial dimension. We will show below that motor-driven buckling-induced node aggregation (at not too low $P_c$) extends the static percolation limit to a lower value; as long as the effective coordination number $z_\textrm{eff}$ of the aggregates satisfies $z_\textrm{eff}>z_c$, global force balance can be achieved and stable patterns emerge.

At the isostatic point, $z_c=4$ in 2D, the shear modulus of an elastic network vanishes. While here, either for the disordered network which undergoes an arrested phase separation (Fig.~\ref{percolating}b left), or for the regular network comprised of tense filaments or bundles (Fig.~\ref{percolating}b middle), the network maintains a finite shear modulus even below $z_c$.
This unexpected situation arises because of the asymmetry in the elastic response of individual filaments to stretching and to compression; stretching stiffens the bonds whereas compressive load is relieved through filament buckling.
At $z<z_c$, local force imbalance at individual nodes is sensed and mitigated (or enhanced) by susceptible (or load-resisting) motors which tend to displace the node along (or against) the force acting on it. Consequently, a considerable fraction of the bonds are buckled while the neighboring ones are stretched. This yields loose ($s>0$) or tight ($s<0$) node aggregates which can be thought of as the vertices of an effective network composed of tense filaments only. This effective network having a smaller number of ``nodes" satisfies $z_\textrm{eff}>z_c$, thus maintaining a stable percolating structure.
This apparent extension to a lower static percolation limit stems from the fact that a cluster of nodes connected by buckled bonds forms an effective vertex which may well have a higher connectivity than the original individual nodes. As shown in the schematic (Fig.~\ref{percolating}d), as the central bond buckles under motor action, the original system having two nodes with $z=3$ which is unstable becomes one having a single vertex (marked by a dashed circle) with $z=4$ and thus being marginally stable. Therefore, such motor-driven buckling-induced node aggregation provides a mechanism to stabilize the network structure below the apparent static percolation limit.
This dynamic mechanism represents a related yet different perspective from that of an earlier study \cite{Chase_pull out soft}, where the soft modes being pulled out originate from bending degrees of freedom and the tensing of the network is due to stretching of the athermal stiff filaments without changing the effective coordination number.
A transparent illustration of this mechanism can be seen for mechanically stable structures formed by load-resisting motors (Fig.~\ref{percolating}b middle), whereby each vertex concentrates more than one nodes which are connected by completely buckled filaments (note that vanishing excluded volume allows overlapping between the nodes). The resulting network sitting on a triangular-like lattice consists of tense filaments only, with an effective coordination number $z_\mathrm{eff}\ge z_c=4$.

%\subsubsection{Dynamic development of the tense regular structure}

\begin{figure}[htb]
\centerline{\includegraphics[angle=0, scale=0.45]{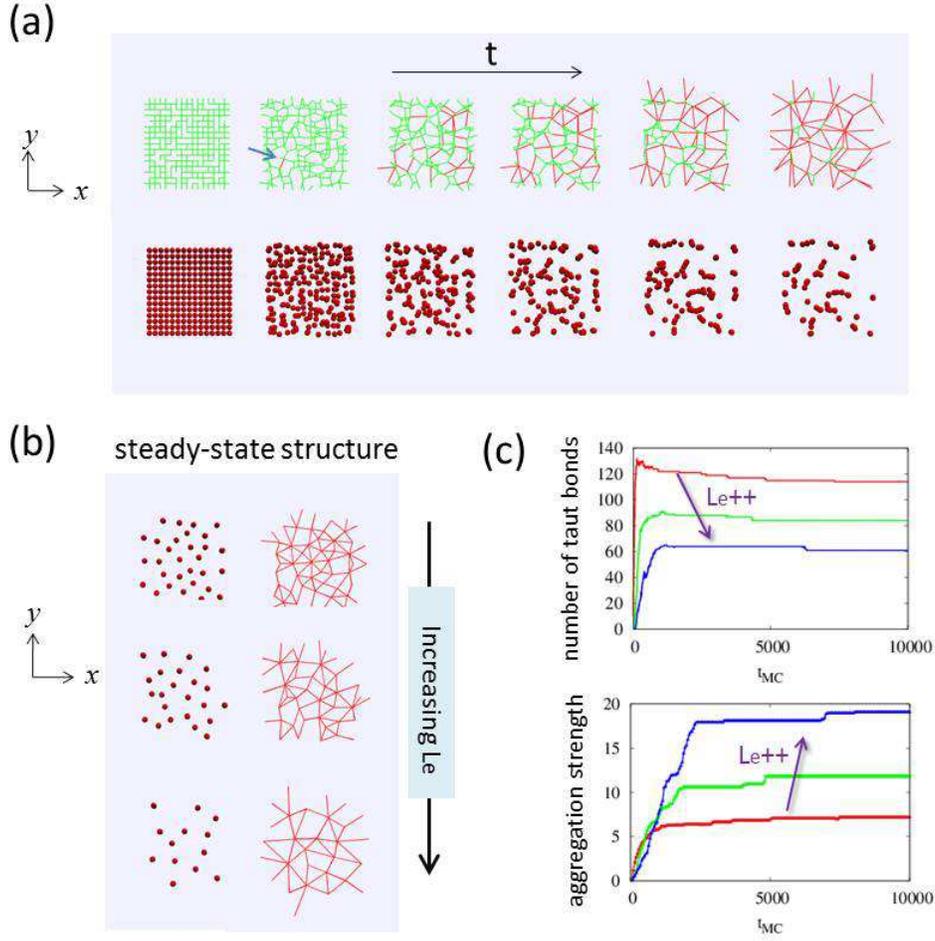}}
\caption{Development of tense regular structures and the role of elastic nonlinearity.
(a) Snapshots during the ``nucleation" of bond-stretching events. Arrow in the second-column upper panel indicates the first stretched filament (red). Buckled filaments (green) are concentrated at the junctions of the coarsening network. (b) Steady-state structures for $L_e=1.2, 1.5$ and $1.8$ (top to bottom). (c) Time courses of the number of stretched bonds (upper) and the aggregation strength (lower) defined as the height of the innermost peak of the pair distribution function of the nodes. The model parameters are $s=-0.5, \beta\gamma=5, l=0.05, \kappa=0.1$ and $P_c=0.8$.}
\label{stretching_nucleation}
\end{figure}

We finally demonstrate the dynamic development of such tense regular structures (Fig.~\ref{stretching_nucleation}) and show that the overall effective connectivity of the network also depends on elastic nonlinearity, in particular the relaxed length, of individual filaments.
Force-sensitive motor processes can capture instantaneous force imbalance at individual nodes and ``nucleate" bond-stretching events thereafter. (Arrow in Fig.~\ref{stretching_nucleation}a indicates the emergence of the first stretched bond.) The structure coarsens as more node aggregates form and more bonds become stretched, in approach to the steady-state structure consisting of tense asters.
The characteristics of the steady-state structure depends on the nonlinearity in the elastic response, characterized by the relaxed length $L_e$ of individual filaments. As $L_e$ increases, it takes a longer time for the network to relax to the steady state (Fig.~\ref{stretching_nucleation}c) , i.e. to achieve a global force balance. On the other hand, a broader range of buckling behavior due to a longer $L_e$ allows more freedom for the system to explore the configurational space before getting arrested globally.
This leads to a higher degree of node aggregation (Fig.~\ref{stretching_nucleation}c lower, Fig.~\ref{stretching_nucleation}b left column) and thus stronger stretching of a smaller number of taut bonds (Fig.~\ref{stretching_nucleation}c upper, Fig.~\ref{stretching_nucleation}b right column).
This dependence suggests that it might be possible to control the effective connectivity of the whole structure by manipulating the nonlinearity of individual filaments, for instance by adding bundling proteins that shorten the relaxed length.

\subsection{Directed transport by an anchored isotropically contracting 3D network}

\begin{figure}[htb]
\centerline{\includegraphics[angle=0, scale=0.6]{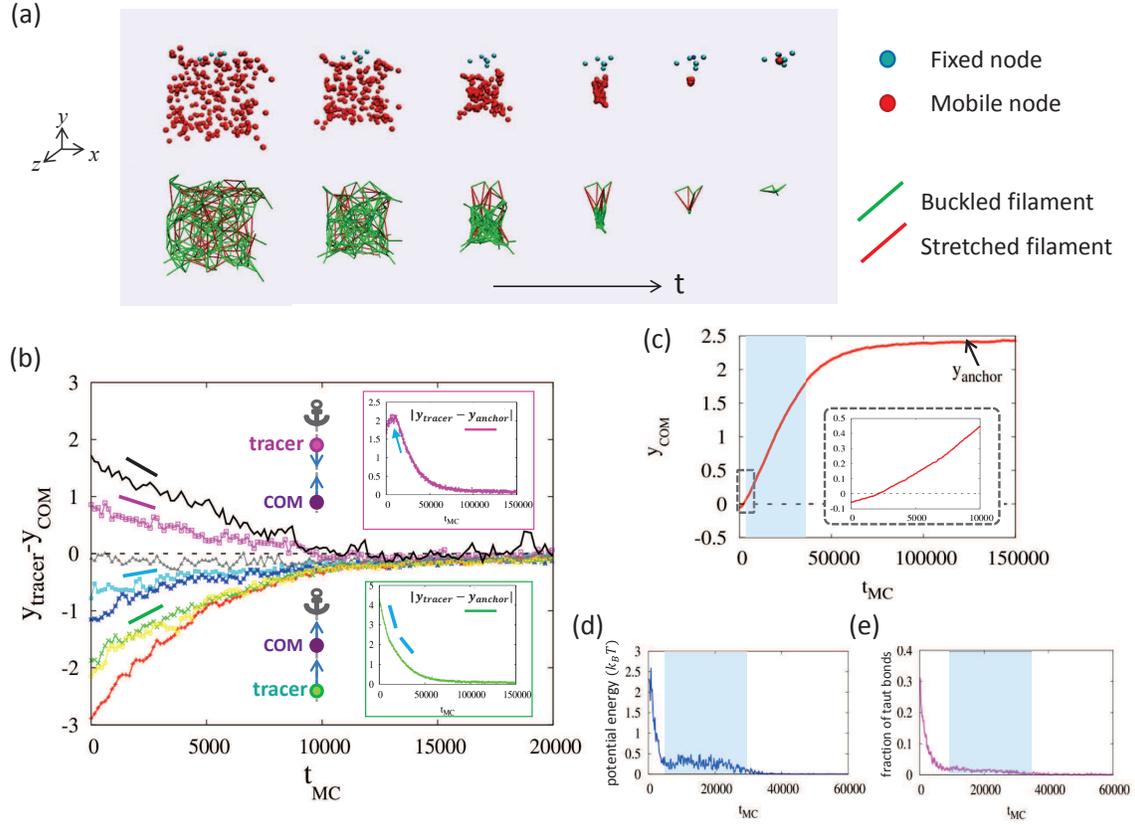}}
\caption{Transport by an anchored isotropically contracting model network.
(a) Snapshots during the isotropic contraction and translocation toward the anchor (cyan spheres).
(b) Time course of the position of the tracer nodes with respect to the center of mass (COM) of the finite network. Insets show the time evolution of the distance from the anchor for two tracer nodes, one above the COM (upper, magenta) and the other below (lower, green). Corresponding schematics indicate the direction of motion for the tracer and the COM. (c) COM motion toward the anchor. Inset shows the slow motion at short times.
(d) Time courses of the potential energy and the corresponding fraction of taut bonds. Shaded regions in (c-e) indicate the linear regime where the COM moves toward the anchor at a constant speed.
The model parameters are $L_e=1.2, \beta\gamma=5, l=0.05, \kappa=0.01, P_c=0.5$ and $s=1$.}
\label{transport}
\end{figure}

To investigate the dynamic evolution of an anchored actomyosin network, mimicking the experimental setting for the F-actin driven chromosome transport in starfish oocyte, \cite{intracellular transport} we perform similar stochastic simulations as those for the 2D case, but now on a 3D finite amorphous network with free boundaries. Again the nodes interact with each other through the cat's cradle type potential and are subject to anti-correlated kicks driven by myosin motors. Several nodes at the center of the top layer of the network are fixed (cyan spheres in Fig.~\ref{transport}a upper row), mimicking the mechanical anchoring to certain cellular structures such as the cell cortex.

Fig.~\ref{transport}a shows the snapshots of node configurations (upper) and the corresponding bond structures (lower) during the process of network contraction and translocation. It appears that homogeneously distributed contractile activity over the percolating network drives isotropic contraction toward the center of mass (COM) that moves toward the mechanical anchor.

To quantify the motion, we show in the main plot of Fig.~\ref{transport}b the anchor-ward projection, i.e. the $y$ component, of the coordinate for several representative tracer nodes. Regardless of their locations in the network, nodes with similar initial distance from the COM exhibit similar speed toward it (the average slope indicated by the colored segments), indicating that the contraction is isotropic. Much as in cosmological models (Hubble's law), moreover, this speed depends almost linearly on the initial distance from the COM, suggesting a homogeneous contraction toward the COM.

Insets of Fig.~\ref{transport}b show the trajectories of two tracer nodes with respect to the anchor, $|y_{tracer}-y_{anchor}|$, over a long time course.
The node above the COM (upper inset) is first pulled away from the anchor due to isotropic contraction toward the COM (ascending part of the curve), followed by the contracted network moving toward the anchor as a whole (descending part). For the node below the COM (lower inset), in addition to the COM motion toward the anchor, an extra velocity component in the same direction arises from contraction toward the COM, leading to a faster approach to the anchor at early times (a steeper slope compared to that at later times).
We present the trajectory of the COM motion in Fig.~\ref{transport}c. On short time scales (inset) when isotropic contraction dominates, COM moves slowly toward the anchor located at $y_{anchor}$ (indicated by arrow). During intermediate times (shaded region) the COM moves at a constant speed toward the anchor, as seen experimentally.\cite{intracellular transport} In this linear regime, the fraction of taut bonds almost remains the same (Fig.~\ref{transport}e) and the potential energy due to bond stretching (Fig.~\ref{transport}d) fluctuates about a constant mean value; these are consistent with the observation that the COM moves under tension due to several stretched bonds that connect the contracted network to the anchor. At longer times, these remaining tense bonds also contract, pulling the network to the anchor eventually (see final snapshots in Fig.~\ref{transport}a). The COM motion stops once all the bonds have buckled.

Our simulations suggest that homogeneously distributed anti-correlated motion in an anchored network of nonlinear elastic filaments indeed captures many of the features of the transport behavior observed experimentally. This finding supports the idea of an essential role of myosin motors in correlating local movements of the nodes and their interplay with network architecture to drive coordinated motion of the entire system. More intriguingly, it suggests a novel intracellular transport mechanism in which contractile subunits organize into three dimensional networks to drive global contraction; mechanical anchoring to cellular structures then directs such contractile motion to distinct subcellular locations.

\section{Conclusion and Outlook}

In sum, we have constructed a minimal microscopic model of an actomyosin network that incorporates essential elements for self-organization. This dynamic model not only captures active pattern formation in large structures, but also reveals the key physical ingredients for directed transport by a finite motor-filament network.
We first show that the non-trivial force dependence of motor kinetics is essential for the formation of mechanically stable structures; without mechanochemical coupling (i.e. $s=0$), motor activity would destabilize the structure leading to a complete collapse.
In percolating networks driven by susceptible motors, buckling instability, or more generally, asymmetry in the elastic response of individual filaments to stretching and to compression, is shown to allow localization of collapse events, thus keeping the overall structure functional despite there being local mechanical failure.
We then demonstrate that motor-driven buckling-induced node aggregation provides a dynamic mechanism that stabilizes patterns below the static percolation threshold. Effective connectivity of networks stabilized in this fashion depends on elastic nonlinearity of individual filaments, characterized by their relaxed length. We also find that, in a well-connected structure, coordinated motor action could suppress thermal noise on large time scales, 2D configurations thus remaining planar regardless of motor susceptibility. Our simulations on a finite anchored actomyosin network suggest that the same molecular machinery, responsible for cytokinesis and morphogenesis based on macroscopic contraction in low dimensions, may also give rise to 3D intracellular directed transport: Efficient translocation of organelles can be achieved by isotropic contraction of a 3D amorphous actomyosin network towards a moving center.

Despite being constructed in the context of cytoskeletal networks, the present model might have implications for other systems having nonlinear interactions and mechanochemical coupling, since motors and nodes can represent different objects in different circumstances. For instance, the mechanism of active contractility in actomyosin networks is likely to work also for blood clots, where platelets play a similar role to myosin motors and exert internal stresses. Similar principle may still apply yet on larger scales to other tissues.

One interesting aspect absent from our current study is the transient action of crosslinking proteins and the unbinding kinetics of motor proteins from the filaments, which may allow more efficient remodeling of living cells and lead to rich dynamical behavior described by several characteristic time scales. By introducing load-dependent bond rupture rate and motor detachment rate, we may study frequency-dependent rheological properties of the cytoskeletal network, drawing connections to studies using continuum theories.\cite{self organization theory 2, self organization theory 3, self organization theory 4} Such rheological aspects include how the motor kinetics affects the nonlinear viscoelastic responses as well as the low-frequency stress fluctuations, and whether the balance between active sliding, passive crosslinking and forced unbinding can result in a nonequilibrium steady state exhibiting quasi-steady patterns.\cite{steady clusters chap1, scale-free clusters}

In an attempt to characterize the deviation from thermal equilibrium behavior due to the energy-consuming motor processes, we have previously studied several closely related models.
In the limit of small motor steps, we have derived explicit expressions of the effective temperature, $T_\mathrm{eff}$, and the effective diffusion coefficient, $D_\mathrm{eff}$,  which respectively characterize the magnitude of dissipation and fluctuation in the presence of active motor action.\cite{Teff} The predicted relationship between $T_\mathrm{eff}$ and $D_\mathrm{eff}$ as well as their dependence on the motor properties agrees nicely with simulations.
We then went beyond the effective equilibrium regime showing that, at a larger motor step size, onset of dynamic instabilities would give rise to a flow transition where spontaneous collective motion emerges.\cite{flow}
To deal with situations even farther from equilibrium, it might be necessary to consider activated events across energy barriers associated with the mechanical and structural constraints in re-organizing the crowded interior of the cell. As a first step, we have developed a mean-field theory \cite{glassy_network} to examine how the motor characteristics modulate the glassy dynamics observed in osmotically compressed cells. This microscopic theory is based on an analogy between structural glass-forming liquids and random-field Ising magnets, which lies in the heart of the random first order transition theory of the glass transition.\cite{RFOT}
To go further, it would be useful to map the microscopic network model to the framework of aging theory \cite{aging theory}---to quantify how the activated events govern the slow relaxation of the cytoskeletal mechanics.

%with a scale-free size distribution of disjoint clusters as observed in recent experiments.\cite{scale-free clusters}

Moreover, we have focused on an effectively one-component system composed of implicitly motor-attached and motor-free filaments. Future work will aim to model an explicit two-filament system, where the myosin motors that slide pairs of actin filaments relative to each other would serve as mobile crosslinks. This additional dynamic element should lead to richer self-organization behavior that better captures the realistic scenario.

%\textbf{- hydrodynamic interactions and flows}

Finally, we have shown earlier \cite{flow} using variational stability analysis combined with computer simulations that, collective sustained flow may arise from cooperative action of force-sensing motors in a well-connected structure. Therein the presence of physical connectivity and force-sensitive motor action provides rigidity of the structure; breaking of continuous translational symmetry underlies the collective flow which also violates time reversal symmetry.
On the other hand, the long-range hydrodynamic coupling mediated by the fluid medium, not yet included in our model, might provide an alternative mechanism for the flow transition, as suggested within continuum frameworks. \cite{Marchetti_HD_PNAS} A linear stability analysis similar to ours has recently been carried out by Woodhouse and Goldstein \cite{Woodhouse_Goldstein_PRL} for a dilute suspension of ``pushers" (short rigid filaments exerting extensile dipolar forces on the fluid) interacting via hydrodynamics only. When confined in a no-slip sphere, this suspension is shown to exhibit swirling patterns above a threshold motor activity.
It will be interesting to examine within our model whether purely hydrodynamic coupling between active elements is also sufficient to generate large-scale coherent flow in motorized models, and to estimate the flow speed by constructing a self-consistent flow field.

%\textbf{- rupture and unbinding kinetics}

\begin{acknowledgments}
We gratefully acknowledge the financial support from the Center for Theoretical Biological Physics sponsored by
the National Science Foundation via Grant PHY-0822283 and the additional support from the Bullard-Welch Chair at
Rice University.

%I am very happy to dedicate this paper to Peter Wolynes, my PhD advisor who inspires me with unique ideas and encourages me to discover myself. I am deeply grateful for his unfailing support over the years.
\end{acknowledgments}

\newpage

%\newpage

\end{document}